\documentclass[aps,prl,twocolumn,showpacs]{revtex4}
\usepackage{bm}
\usepackage{graphicx}
\usepackage{amsmath}
\usepackage{eufrak}
\usepackage{color}
\newcommand{\nix}[1]{}
\begin{document}

\title{Spin currents in diluted magnetic semiconductors (extended version)}

\author{S.D.~Ganichev,$^{1}$ S.A. Tarasenko,$^2$ V.V. Bel'kov,$^2$ P. Olbrich,$^1$  W. Eder,$^1$
D.R.~Yakovlev,$^{2,3}$ V. Kolkovsky,$^4$ W. Zaleszczyk,$^4$ G.~Karczewski,$^4$ T. Wojtowicz,$^4$ and D. Weiss$^1$
}

\affiliation{$^1$  Terahertz Center, University of Regensburg, 93040
Regensburg, Germany}
\affiliation{$^2$A.F.~Ioffe Physico-Technical Institute, Russian
Academy of Sciences, 194021 St.~Petersburg, Russia}
\affiliation{$^3$ Experimental Physics 2, TU Dortmund University, 44221 Dortmund, Germany,}

\affiliation{$^4$ Institute of  Physics, Polish Academy
of Sciences,  02-668 Warsaw, Poland}

\begin{abstract}
Spin currents resulting in the zero-bias spin separation
have been observed in unbiased diluted magnetic
semiconductor structures (Cd,Mn)Te/(Cd,Mg)Te.
The pure spin current generated due to  the electron gas
heating by terahertz radiation is converted into
a net electric current by application of
an external magnetic field. We demonstrate that polarization of
the magnetic ion system
enhances drastically the conversion due to the spin-dependent
scattering by localized Mn$^{2+}$ ions and the
giant Zeeman splitting.
\end{abstract}

\date{\today}

\maketitle

The generation of  spin currents in low-dimensional semiconductor structures
recently attracted a  great  attention.
Pure spin currents represent equal and oppositely directed flows
of spin-up and spin-down electrons. By that the net electric current is zero.
Spin currents lead to a spatial spin separation
and, consequently, an accumulation of the oppositely oriented spins at the  edges of the sample.
Spin currents in semiconductors can be generated by an
electric field, like in the spin Hall effect (for review
see~\cite{spinbook,Fabian}) as well as  by
optical means under interband
optical transitions in non-centrosymmetric bulk and
low-dimensional semiconductors~\cite{Bhat,Tarasenko05p292,Zhao}.
They can also be achieved
as a result of zero-bias spin separation, e.g. by
electron gas heating  followed by spin-dependent energy relaxation
of
carriers \cite{Ganichev06zerobias}. So far in low-dimensional
semiconductors pure spin currents have been reported for
non-magnetic  structures only. However, spin-dependent effects can
be greatly enhanced by exchange interaction of electrons with
magnetic ions in diluted magnetic semiconductors (DMS) like, e.g.
(Cd,Mn)Te\,\cite{spinbook,Dietl,Fur88,Gorini}. Moreover, the strength of
these effects can be widely tuned by temperature, magnetic field
and concentration of the magnetic ions.

Here we report on the observation of the zero-bias spin separation
in (Cd,Mn)Te/(Cd,Mg)Te QWs with Mn$^{2+}$ magnetic ions.
We demonstrate that absorption of terahertz (THz) radiation leads
to  a pure spin current. The effect is investigated
in an external magnetic field converting the
spin separation into a net electric current.
The application of a magnetic field to DMS structures results in the giant
Zeeman spin splitting as well as in the spin-dependent exchange
scattering of electrons by magnetic impurities. Both effects
disturb the balance of the oppositely directed spin-polarized
flows of electrons yielding an electric current. We demonstrate
that the spin-dependent exchange scattering of electrons by magnetic
impurities plays an important role in the current generation
providing a handle to manipulate the spin-polarized currents.

\begin{figure}[t]
\includegraphics[width=0.8\linewidth]{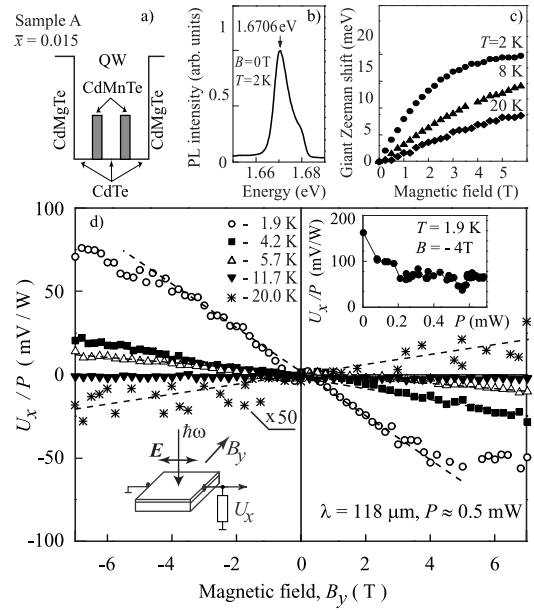}
\caption{(Cd,Mn)Te/(Cd,Mg)Te QW with $\bar{x} = 0.015$ (sample~A).
(a)~Sketch of the structure. (b) Photoluminescence spectrum. (c)
Shift of the PL line corresponding to 1/2 of the total giant
Zeeman splitting. (d)~Magnetic field dependence of the voltage
signal $U_x$ normalized by the radiation power $P$ in response to
a low excitation power applying $cw$ laser. Insets show the
experimental geometry (left, bottom) and $U_x/P$ as a function of
the radiation power obtained at $B=-4$~T (right, up). }
\label{figure1_Bfield_010408AIII}
\end{figure}

We study spin currents on (Cd,Mn)Te/(Cd,Mg)Te) single
QW structures grown by molecular beam epitaxy on (001)-oriented
GaAs substrates~\cite{Crooker, Egues,Jaroszynski2002}. The quantum wells were grown
by a digital alloy technique \cite{Kneip2006}, i.e. by inserting
evenly spaced thin layers of Cd$_{1-x}$Mn$_x$Te during the growth
of 10~nm wide CdTe-based QW [Fig.~\ref{figure1_Bfield_010408AIII}(a)].
Two  DMS samples were fabricated: Sample~A with two insertions of
Cd$_{0.8}$Mn$_{0.2}$Te and a nominal thickness of three monolayers
and sample~B with three single monolayers of
Cd$_{0.86}$Mn$_{0.14}$Te. The samples were modulation doped with
Iodine donors introduced in a top Cd$_{0.76}$Mg$_{0.24}$Te barrier
in a 15~nm distance from the QW.
As a reference structure we use sample~C without Mn insertion.
The samples are characterized by transport measurements and optical
spectroscopy. The photoluminescence (PL) spectrum of sample~A
shown in Fig.~\ref{figure1_Bfield_010408AIII}(b) is typical for
modulation-doped QWs~\cite{Kel03}.
The linewidth of 11~meV corresponds to the  Fermi energy,
$E_F$, and is in good agreement with the transport
measurements. In external magnetic fields this line shows a strong
spectral shift to low energies, reflecting the giant Zeeman
splitting of the band states in DMS~\cite{Fur88}. This shift, shown
in Fig.~\ref{figure1_Bfield_010408AIII}(c), is strongly temperature
dependent and is about 2.5 times
larger than the giant Zeeman splitting of the
conduction band states.
The samples parameters are summarized in Table~\ref{table1}.
For the photocurrent experiments square shaped ($5\times5$~mm$^2$)
specimens with a pair of ohmic contacts centered on the sample
edges along the direction $x
\parallel [1\bar{1}0]$ were prepared
[see inset in Fig.~\ref{figure1_Bfield_010408AIII}(d)].

\begin{table}
\caption{\label{table1} Sample parameters. Here electron mobility, $\mu$, and electron concentration,
$n_e$, in the QW was evaluated from transport experiments and
$\bar{x}$ is the effective average
concentration of Mn, determining the electron spin splitting,
which was estimated from the giant Zeeman shift of interband emission line
[Fig.~\protect \ref{figure1_Bfield_010408AIII}(c)].
}
%
\begin{tabular}{cccccc}
\hline
sample & \,\,\,\,
$x$ & \,\,\,\,  $\bar{x}$ &\,\,\,\, $\mu$ at 4.2~K & \,\,\,\, $n_e$ at 4.2~K & \,\,\,\, $E_F $\\

       & \,\,
                       &                     & \,\, cm$^2$/Vs& 10$^{11}$ cm$^{-2}$ & \,\, meV\\
\hline

A 

                     &         0.20  &            0.015    &             9500       &  4.7  & 11.7 \\

B 

                     &         0.14  &            0.013    &             16000       &  6.2  & 15.4 \\

C %

                      &         0  &            0    &             59000       &  4.2  & 10.4 \\

       \hline
\end{tabular}
\end{table}

To generate spin photocurrents we heat an electron gas
applying  radiation of a low power continuous-wave
($cw$)
optically pumped
CH$_3$OH laser operating at a wavelength
$\lambda$~=~118~$\mu$m and a power $P \approx
$~0.5~mW reaching the sample. In addition we used a high power pulsed
NH$_3$ laser
providing 100~ns pulses with $\lambda$~=~148~$\mu$m
and  $P$ up to 40~kW~\cite{book}.
THz radiation with photon energies  $\hbar \omega \approx 10$~meV is chosen
to induce only free carrier
absorption.
The radiation is linearly polarized
with the polarization vector aligned along the
$x$-axis.
An in-plane magnetic field ${\bm B} \parallel y$  (up to 7~T)
is used to align the Mn$^{2+}$ spins  and to convert the
spin current into a spin-polarized charge current.
The geometry of the experiment is sketched in the
inset of Fig.~\ref{figure1_Bfield_010408AIII}(d).
The photocurrent is generated at normal incidence of radiation
in (001)-oriented unbiased devices and measured perpendicularly
to the magnetic field.
This experimental configuration excludes other effects known
to cause photocurrents~\cite{Ganichev06zerobias}.
The radiation of the $cw$ laser is modulated at $225$~Hz.
A photocurrent is measured across the 1\,M$\Omega$ load resistance
and recorded after 100~times voltage
amplification by a lock-in amplifier.
Whereas, in response to
the pulsed radiation it is measured by the voltage drop
across a 50~$\Omega$ load resistor.

\begin{figure}[t]
\includegraphics[width=0.8\linewidth]{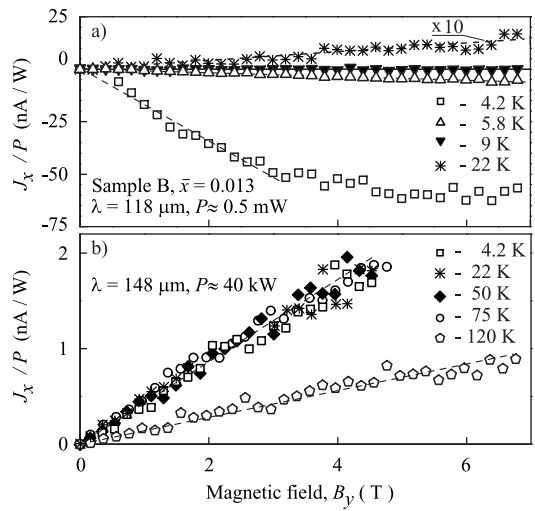}
\caption{Magnetic field dependence of the photocurrent $J_x$
normalized by the radiation power $P$ measured in sample~B  with
$\bar{x}= 0.013$. a) Low power excitation ($P \approx 0.5$~mW)
with radiation of the $cw$ laser b) High power excitation
($P~\approx~40$~kW) with radiation of the pulsed laser~\protect
\cite{footnote}. }
\label{figure3_Bfield_091906A_cw_pulsed}
\end{figure}

Irradiating the DMS sample~A with  low power $cw$ THz
radiation  we observe a voltage signal, $U_x$,  for non-zero
magnetic fields [Fig.~\ref{figure1_Bfield_010408AIII}(d)]. The
signal polarity reverses with the change of the magnetic field
direction. The signal is detected in a temperature  range from 1.9 to 20~K.
As an important result we observe that a
cooling of the sample changes the signal sign and
increases its absolute values by more than two orders of
magnitude. While at moderate temperatures the signal depends linearly on $B$,
at low $T$ we observed a saturation of the photocurrent with rising magnetic field strength
[see the data for $T$~=~1.9~K in Fig.~\ref{figure1_Bfield_010408AIII}(d)].
Similar results are obtained for sample~B [Fig.~\ref{figure3_Bfield_091906A_cw_pulsed}(a)].
Both temperature and magnetic field dependences
are typical for magnetization
of DMS due to polarization of the Mn spin system in an
external magnetic field.

\begin{figure}[t] 
\includegraphics[width=0.8\linewidth]{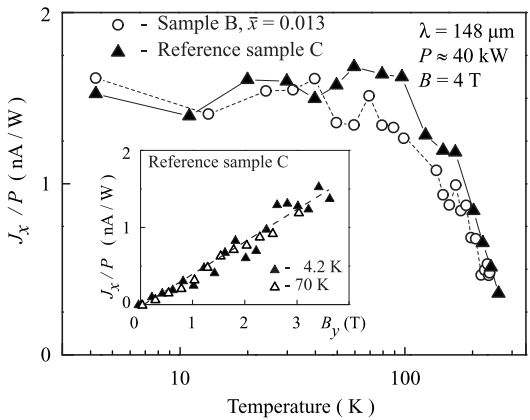}
\caption{Temperature dependence of the normalized photocurrent
$J_x/P$ excited by the radiation of $\lambda~=$~148~$\mu$m at high
excitation power of about 40~kW. Results are plotted for the sample~B
($\bar{x}= 0.013$)
and the
reference sample~C. The inset shows the magnetic field dependence
of the photocurrent for sample~C.
}
 \label{figure4_temperature_pulsed}
\end{figure}

Our experiments demonstrate that THz radiation induced
photocurrent in DMS  is controlled by the exchange interaction of
electrons with Mn$^{2+}$ ions. The well-known effect of the
exchange interaction is the giant Zeeman splitting~\cite{Fur88},
also detected in our samples by PL measurements [see
Fig.~\ref{figure1_Bfield_010408AIII}(c)]. At low
temperature\textbf{s} the exchange spin splitting overcomes the
intrinsic one. The  energy separation of the spin-up and spin-down
electron subbands in (Cd,Mn)Te reads \cite{Fur88,Kneip2006}
\begin{equation}
\label{eq2}
E_{Z}(B) = g_e \mu _B B +  \bar{x} S_0 N_0 \alpha {\rm B}_{5/2}
\left( {\frac{5 \mu _B g_{Mn} B}{2 k_B (T_{Mn} + T_0 )}} \right) \, ,
\end{equation}
where $k_B $ is the Boltzmann constant and  $\mu _B$  the Bohr
magneton. Here the first term stands for the intrinsic spin
splitting with the electron $g$-factor ($g_e=
-1.64$~\cite{Sir97}).  The second term is caused by exchange and
contributed by the Mn$^{2+}$ $g$-factor ($g_{Mn} = 2$) and the
temperature of the Mn-spin system $T_{Mn}$. Phenomenological
parameters $S_0$ and $T_0$ allow one to account for the Mn-Mn
antiferromagnetic interactions within the magnetic ion system.
$\rm{B}_{5/2}\left( \xi \right)$ is the modified Brillouin
function of the argument in the brackets, $N_0 \alpha = 220$~meV
is the exchange integral for conduction band electrons and $N_0$
is the number of cations per unit volume.

It follows from Eq.~(\ref{eq2}) that the Zeeman splitting
has a strong temperature dependence and reverses its sign due to
opposite signs of $g_e$ and $N_0 \alpha$. This explains the
 sign inversion of the photocurrent in
Fig.~\ref{figure1_Bfield_010408AIII}(d). However, while the
photocurrent at $T$~=~20~K  already changes its direction compared
to lower temperatures [Fig.~\ref{figure1_Bfield_010408AIII}(d)]
the sign of the Zeeman splitting, detected by PL,  remains the same
[Fig.~\ref{figure1_Bfield_010408AIII}(c)]. We attributed this fact
to the heating of the Mn$^{2+}$ spin system over
the lattice temperature. Such an effect has been reported for
DMS~\cite{Kel02}.
To check this assumption we investigate the power dependence of
the signal. The inset of Fig.~\ref{figure1_Bfield_010408AIII}(d)
demonstrates that at higher power the normalized signal decreases,
indicating
a reduction of the exchange effects due to an increase of
$T_{Mn}$. To make the effect of heating more pronounced we applied
the radiation of a pulsed THz  laser with an eight orders of
magnitude higher power than that of the $cw$ laser. The data at low
power and high power excitations are shown in
Fig.~\ref{figure3_Bfield_091906A_cw_pulsed} for sample~B. We
observe that the substantial increase of the radiation power
results in the change of the signal polarity at $T~<~15$~K.
Moreover, the drastic temperature dependence detected at low power
excitation (Fig.~\ref{figure3_Bfield_091906A_cw_pulsed}(a))
in the range from 4.2 to 22~K  disappears and
the photocurrent becomes almost independent of the sample temperature
(Fig.~\ref{figure3_Bfield_091906A_cw_pulsed}(b)).
The $T$-dependence under high power
excitation is shown in Fig.~\ref{figure4_temperature_pulsed}. The
signal is about constant for $T~<~100$~K and at higher
temperatures it decreases with rising $T$.
The signals in our pulse measurements are substantially higher than 
that at low power. 
%
As a result we
obtain measurable signals at higher $T$ as well as in the
reference non-magnetic sample~C
(Fig.~\ref{figure4_temperature_pulsed}). It is remarkable that at
high power excitation absolute values and the temperature
dependences of the DMS sample~B and
the reference sample~C are very similar to each other.
This fact indicates that the presence of Mn$^{2+}$ ions does not
contribute to the current at this experimental conditions (high
power excitation). Note, that the temperature dependence in
Fig.~\ref{figure4_temperature_pulsed} is also similar to that
previously reported for non-magnetic semiconductors~\cite{Ganichev06zerobias}.

\begin{figure}  
\includegraphics[width=0.6\linewidth]{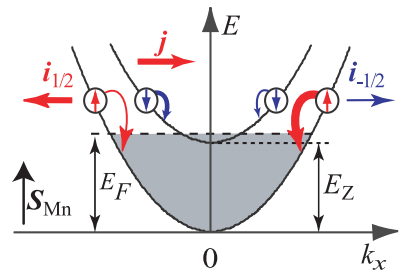}
\caption{Model of the zero-bias spin separation and conversion of
the pure spin current into a net electric current due to an
external magnetic field.  The zero-bias spin separation is caused by
the scattering assisted energy relaxation of the electron
gas heated by radiation. The spin dependent part of the scattering matrix
linear in ${\bm k}$ and ${\bm \sigma}$
results in an asymmetric relaxation:
transitions to positive and negative $k^\prime_x$-states occur with different
probabilities. This is indicated by bent arrows of different
thickness. Thus, the transitions
yield an asymmetric occupation of both subbands and hence electron flows.
Without magnetic field these flows are oppositely directed and equal to each other.
The Zeeman spin splitting causes a preferentially occupation of
one subband (the spin-up states in the figure) and disturbs the balance.
As a result the pure spin current is converted into a spin polarized electric current.
In DMS at low temperatures the conversion
is amplified by  the giant Zeeman spin
splitting and the spin-dependent
electron scattering by polarized magnetic ions.
If instead of the spin-up subband the spin-down subband is
preferentially occupied the current direction is reversed.
}
\label{Figure5model}
\end{figure}

We now turn to  microscopic mechanisms responsible for photocurrent generation.
In case of Drude absorption, photocurrents stem from spin-dependent asymmetry 
of the optical transitions accompanied by scattering and/or from energy relaxation\,\cite{Ganichev06zerobias}. 
Here we focus on the energy relaxation of electron gas heated by THz yielding 
a polarization independent photocurrent.
%
While the first mechanism depends
on the radiation polarization the latter one is polarization independent.
Investigating photocurrent  in DMS at low temperature we do not
observe any polarization dependence.
Therefore, the spin separation mechanism of interest here is based on the
electron heating by THz radiation followed by energy relaxation.
Usually, energy relaxation is considered to be spin-independent.
However, in gyrotropic media, like CdTe- and (Cd,Mn)Te-based QWs
investigated here, the spin-orbit interaction adds an asymmetric
term to the scattering matrix element. This term is proportional
to $\sigma_{\alpha}(k_{\beta}+k^\prime_{\beta})$,
 where $\bm{\sigma}$ is
the vector composed of the Pauli matrices, $\bm{k}$ and $\bm{k}^\prime$ are the
initial and scattered electron wave vectors. Therefore, energy relaxation
processes became spin-dependent. This is indicated for both spin subbands by
bent arrows of different thicknesses in Fig.~\ref{Figure5model}. The asymmetry
of the electron-phonon interaction results in the spin current: The oppositely
directed electron fluxes in two spin subbands, $\bm{i}_{\pm1/2}$, are of equal
strength and the total charge current is zero, $\bm{j} = e
(\bm{i}_{+1/2}+\bm{i}_{-1/2})=0$. Here $e$ is the electron charge.
Nevertheless, a finite pure spin current $\bm{J}_{s} = (\bm{i}_{+1/2} -
\bm{i}_{-1/2})/2$ is generated. This leads to a spatial spin separation and
spin accumulation at the sample edges.

The application of an external magnetic field introduces an imbalance between
the fluxes $\bm{i}_{\pm1/2}$ giving rise to a net electric current
$\bm{j}=e(\bm{i}_{+1/2}+\bm{i}_{-1/2})$. An obvious mechanism of the magnetic
field induced imbalance of the oppositely directed spin flows, also relevant
for non-magnetic semiconductors, is related to the Zeeman splitting of electron
states~\cite{Ganichev06zerobias}. This process is sketched in
Fig.~\ref{Figure5model}. Indeed, the fluxes $\bm{i}_{\pm1/2}$ depend on the
free carrier densities in the spin-up and spin-down subbands, $n_{\pm1/2}$.
Therefore, in a Zeeman spin-polarized system, where $n_{+1/2} \neq n_{-1/2}$,
the fluxes $\bm{i}_{\pm1/2}$  do no longer compensate each other yielding a net
electric current. In DMS the electron Zeeman splitting is giantly enhanced due
to the exchange interaction between free electrons and Mn$^{2+}$  ions. In the
case of a low electron spin polarization, the equilibrium electron spin per
electron is given by $s = - E_Z/(4\bar{E})$ and the net current caused by the
Zeeman splitting is given by~(see~Appendix~I)
\begin{equation}\label{current1}
\bm{j}_{Z} = - 4 e \frac{E_{Z}}{4 \bar{E}} \left( n_e \frac{\partial
\bm{J}_{s}}{\partial n_e} \right) \:,
\end{equation}
where $\bar{E}$ is a characteristic electron energy, which is equal to $E_F$
for the degenerated electron gas and $k_B T$ for the non-degenerated gas. The
spin current is considered here as a function of the carrier density $n_e$. In
particular, for the Boltzmann statistics, where $\bm{J}_{s} \propto n_e$ and
$n_e
\partial \bm{J}_{s} / \partial n_e =\bm{J}_{s}$, Eq.~(\ref{current1})
yields
\begin{equation}\label{current1b}
\bm{j}_{Z} = - e \frac{E_{Z}}{k_B T}\bm{J}_{s} \:,
\end{equation}
For the Fermi distribution, the derivative $\partial \bm{J}_{s} /
\partial n_e$ vanishes if the spin current is caused solely by
$\bm{k}$-linear terms in the matrix element of the electron-phonon
interaction and is non-zero if higher order in $\bm{k}$ terms
contribute to the spin current.

Equations~(\ref{current1}) and~(\ref{current1b}), showing that
$j_Z \propto E_Z$,  explain together with Eq.~(\ref{eq2}) the most
striking experimental result: The sign inversion of the photocurrent with decreasing
temperature. Indeed, at high temperatures
the  last term in Eq.~(\ref{eq2}) vanishes and
only the intrinsic Zeeman splitting
is responsible for the effect. The
temperature decrease results in
the giant Zeeman splitting whose sign is opposite to the intrinsic one.
Thus, the direction of the photocurrent determined by the sign  $E_Z$ reverses.

This mechanism alone, however, does not explain the quantitative
change of the current magnitude. It follows from the
comparison of the temperature dependences of the  Zeeman
splitting of the electron subbands, estimated
from the PL data and the photocurrents.
Indeed, for sample~B at $B=3$~T and $T=4.2$~K, the value
of the giant Zeeman splitting is 2.6~meV. It exceeds by an order
of magnitude the intrinsic Zeeman splitting of 0.25~meV.
Whereas comparing the data obtained at the same magnetic field for 22~K,
where only the intrinsic effect is present~\cite{footnote},  and 4.2~K
we get that the current strength changes much stronger, by a factor of about 40
(see Fig.~\ref{figure3_Bfield_091906A_cw_pulsed}). This variation
of the signal is caused by Mn-related spin-dependent properties of
DMS only and is much larger than that of the Zeeman
splitting~\cite{footnote2}.

The mechanism, resulting in additional contribution to the
conversion of the spin separation into the net current, is
specific for DMS and is caused by the well known spin-dependent
electron scattering by polarized magnetic ions~\cite{Fur88}. In
external magnetic fields, when the Mn ions are spin polarized, the
scattering rate of electrons with the spins aligned parallel and
antiparallel to the Mn spins becomes different~\cite{Egues}. This results in
two different momentum relaxation times, $\tau_{p,+1/2}$ and
$\tau_{p,-1/2}$, in the spin subbands. Since the electron fluxes
$\bm{i}_{\pm1/2}$ are proportional to $\tau_{p,\pm 1/2}$, the
polarization of Mn spins leads to a net electric current,
$\bm{j}_{Sc}$. To obtain $\bm{j}_{Sc}$  we assume that the
momentum relaxation of electrons is governed by their interaction
with  Mn ion localized in QW. The corresponding
Hamiltonian~\cite{Fur88} is given by
\begin{equation}\label{H_el_Mn}
H_{e-Mn}=\sum_i \left[ u - \alpha \,( \bm{S}_i \cdot \bm{\sigma}) \right]
\delta(\bm{r} - \bm{R}_i) \:,
\end{equation}
where $i$ is the Mn ion index, $\bm{S}_i$   the vector composed of the matrices
of the angular momentum $5/2$, $u \delta(\bm{r} - \bm{R}_i)$  the scattering
potential without exchange interaction, $\bm{r}$  the electron coordinate, and
$\bm{R}_i$  the Mn ion position. The electron scattering by the Mn potential
determined by $u$ is usually stronger than the exchange scattering determined
by $\alpha$. Note, that the parameter $\alpha$ in Eq.~(\ref{H_el_Mn}) is also
responsible for the giant Zeeman splitting in Eq.~(\ref{eq2}). Then, for the
case of $|\alpha| \ll |u|$, we derive~(see Appendix~II)
\begin{equation}\label{current2}
\bm{j}_{Sc} = 4 e \frac{\alpha}{u} \, \bm{J}_{s} S_{Mn}  \:,
\end{equation}
where $S_{Mn}$ is the average Mn spin along the magnetic field direction. We
note that at low temperatures when DMS properties dominate the photocurrent
both $\bm{j}_{Sc}$ and  $\bm{j}_Z$ have the same direction because the average
electron spin caused by the Zeeman effect is parallel to $\bm{S}_{Mn}$. The
total electric current is given by the sum of both contributions $\bm{j} =
\bm{j}_{Z} + \bm{j}_{Sc}$. In the case of a full spin polarized electron gas
due to Zeeman effect, which can be achieved in DMS in reasonable magnetic
fields, the electron flow in one of the spin subbands vanishes. Therefore, the
electric current becomes independent of the magnetic field strength and carrier
statistics and is given by  $\bm{j}=\mp 2e \bm{J}_{s}$, where $\mp$ corresponds
to $\pm$ sign of the Zeeman splitting.

To summarize, the electron gas heating in low-dimensional
diluted magnetic semiconductors
results in the generation of the pure spin current and
correspondingly in the zero-bias spin separation.
We show experimentally and theoretically that the
carrier exchange interaction with localized magnetic spins in DMS
giantly amplifies the conversion of the spin current into the electrical
current. Two mechanisms are responsible for that, the giant Zeeman
splitting of the conduction band states and the spin-dependent
carrier scattering by localized Mn spins polarized by an external
magnetic field. In weak magnetic fields for a degenerated electron
gas the scattering mechanism dominates the current conversion.

\acknowledgments We thank L.~V. Litvin, E.~L.~Ivchenko, L.~E.~Golub,
S.~N.~Danilov, N.~S.~Averkiev, and Yu.~G.~Semenov. The financial support from
the DFG, RFBR, Polish Ministry of Science and Higher Education and Foundation
for Polish Science is gratefully acknowledged.

\setcounter{equation}{0}
\renewcommand{\theequation}{A\arabic{equation}}
\begin{appendix}
\subsection{Appendix I}\label{A1}

The electric current caused by different population of the spin-up and
spin-down subbands in the magnetic field is given by
\begin{equation}
\bm{j}_Z = e [\bm{i}_{+1/2}(n_e/2+\delta n ) + \bm{i}_{-1/2}(n_e/2-\delta n)]
\end{equation}
\vspace{-0.7cm}
\[
\mbox{} \: \approx e \left[ \bm{i}_{+1/2}(n_e/2) + \bm{i}_{-1/2}(n_e/2) +
\delta n \frac{\partial (\bm{i}_{+1/2}-\bm{i}_{-1/2})}{\partial (n_e/2)}
\right] \:,
\]
where the fluxes $\bm{i}_{\pm1/2}$ are considered here as functions of the
electron densities in the spin subbands, $n_e$ is the total carrier density,
and $\delta n = s n_e$. Since in zero magnetic field the electric current
vanishes, i.e., $\bm{i}_{+1/2}(n_e/2) = - \bm{i}_{-1/2}(n_e/2)$, the spin
current is defined by $\bm{J}_s(n_e) = 1/2 [\bm{i}_{+1/2}(n_e/2) -
\bm{i}_{-1/2}(n_e/2)]$, and $s = - E_Z/(4\bar{E})$, we derive
\begin{equation}
\bm{j}_{Z} = - e \frac{E_{Z}}{k_B T}\bm{J}_{s} \:.
\end{equation}

\subsection{Appendix II}\label{A2}

The electric current contribution caused by difference in the momentum
relaxation times $\tau_{p,\pm 1/2}$ in the magnetic field is given by
\begin{equation}
\bm{j}_{Sc} = e \left( \bm{i}_{+1/2}^{(0)} \frac{\tau_{p,+1/2}}{\tau_{p}^{(0)}}
+ \bm{i}_{-1/2}^{(0)} \frac{\tau_{p,-1/2}}{\tau_{p}^{(0)}} \right) \:,
\end{equation}
where $\bm{i}_{\pm1/2}^{(0)}$ and $\tau_{p}^{(0)}$ are the electron fluxes and
relaxation time in zero field, respectively. For the electron scattering by
polarized Mn impurities, the times $\tau_{p,\pm 1/2}$ assume the form
\begin{equation}
\frac{\tau_{p,\pm1/2}}{\tau_{p}^{(0)}} =  \frac{ \sum_{J=-5/2}^{+5/2} |u \mp
\alpha J |^2 /\,6}{\sum_{J=-5/2}^{+5/2} |u \mp \alpha J |^2 f_{Mn}(J)} \approx
\left[ 1 \pm 2 \frac{\alpha}{u} S_{Mn}  \right]\:,
\end{equation}
where $f_{Mn}(J)$ is the distribution function of Mn spin projections along the
magnetic field, and $S_{Mn} = \sum_{J=-5/2}^{+5/2} J f_{Mn}(J) $ is the average
Mn spin. Then, one obtains for the electric current
\begin{equation}
\bm{j}_{Sc} = 2 e \frac{\alpha}{u} S_{Mn} (\bm{i}_{+1/2}^{(0)} -
\bm{i}_{-1/2}^{(0)}) \approx 4 e \frac{\alpha}{u} \bm{J}_s S_{Mn} \:.
\end{equation}
\end{appendix}


\begin{thebibliography}{99}
\bibitem{footnote} We note the fact that the slopes of $J_x(B)$
dependences detected for $T$ of about 20~K are the same for low
and high power excitations.
\bibitem{footnote2} Our experiments under conditions where Mn
plays no essential role (reference non-magnetic sample and DMS samples  at
high power excitation) demonstrate that the
photocurrent due to intrinsic effect is almost independent
of temperature for $T < 100$~K (see Fig.~\protect\ref{figure4_temperature_pulsed}).
\end{thebibliography}
\end{document}